\documentclass{new_tlp}
\pdfoutput=1

\usepackage{mytlpnojournal}

\usepackage{xspace}
\usepackage{latexsym}
\usepackage{amssymb}   

\usepackage{ifthen}
\newboolean{commentsaon}         
\setboolean{commentsaon}{true}
  \setboolean{commentsaon}{false}  

\newboolean{commentson} 
\setboolean{commentson}{true}

\newcommand{\comment}[1]
{\ifthenelse{\boolean{commentson}\AND\boolean{commentsaon}}
   {{\par\noindent\mbox{}{\small\blue[ *** #1 ]\par}\noindent\par}}{}}

\newcommand{\commenta}[1]
{\ifthenelse{\boolean{commentsaon}}
   {{\par\noindent\mbox{}{\small\color[rgb]{0, .5, 0}[ *** #1 ]\par}\noindent\par}}{}}

\usepackage{pgf}

\usepackage[yyyymmdd]{datetime}

\renewcommand{\today}{2020-07-17}

\usepackage{color}
\newcommand\blue     {\color{blue}}

\usepackage{hyperref}

\newtheorem{theorem}{Theorem}

\newtheorem{lemma}[theorem]{Lemma}

\newtheorem{definition}[theorem]{Definition}

\newcommand*{\seq}[2][n]  {{#2_{1}, \allowbreak \ldots, \allowbreak #2_{#1}}}

\newcommand*{\HU}{{\ensuremath{\cal{H U}}}\xspace}
\newcommand*{\HB}{{\ensuremath{\cal{H B}}}\xspace}
\newcommand*{\TU}{{\ensuremath{\cal{T U}}}\xspace}
\newcommand*{\TB}{{\ensuremath{\cal{T B}}}\xspace}
\newcommand*{\M}{{\ensuremath{\cal M}}\xspace}
\newcommand*{\Rel}{{\ensuremath{\cal R}}\xspace}

\usepackage[mathscr]{euscript}

\newcommand*{\NN}{{\ensuremath{\mathbb{N}}}\xspace}

\newcommand*{\myunderscore}{\mbox{\tt\symbol{95}}}

\newcommand*{\nqueens}{{\sc nqueens}\xspace}
\newcommand*{\APPEND}{{\sc append}\xspace}

\title{On correctness
  of an $n$ queens program}

\submitted{\today}
\revised{. . .}\accepted{. . .}

\author   {W{\l}odzimierz Drabent%
\vspace{1ex}
\\
{\rm\normalsize\small\footnotesize
 \today%
        {\ifthenelse{\boolean{commentson}\AND\boolean{commentsaon}}
           {\blue\quad  [Version 4.5 with private comments]}{}%
        }%
}
   \\ \\
     \normalsize\small
     \begin{tabular}{c}
          Institute of Computer Science,
          Polish Academy of Sciences,\\
          ul. Jana Kazimierza 5,
          01-248 Warszawa, Poland
          \\ 
         and \\
          Department of Computer and Information Science,
          Link\"oping University\\
          S -- 581\,83   Link\"oping, Sweden      
          \\[.5ex]
          \texttt{drabent\,{\it at}\/\,ipipan\,{\it dot}\/\,waw\,{\it dot}\/\,pl}
     \end{tabular}
}

\begin{document}

\maketitle

\begin{abstract}
 Thom Fr\"uhwirth presented a short, elegant and efficient Prolog program for
 the $n$ queens problem.
 However the program may be seen as rather tricky and one may be not convinced
 about its correctness.   
 This paper explains the program in a declarative way, 
 and provides proofs of its correctness and completeness.
The specification and the proofs are declarative, i.e.\ they abstract from any
operational semantics.  The specification is approximate, it is unnecessary 
to describe the program's semantics exactly.  Despite the program works on
non-ground terms, this work employs the standard semantics, based on
logical consequence and Herbrand interpretations.

Another purpose of the paper is to present an example of precise declarative
reasoning about the semantics of a logic program.

\end{abstract}


\section{Introduction}

 Thom Fr\"uhwirth presented a short, elegant and efficient Prolog program for
 the $n$ queens problem \cite{Fruehwirth91}.
 However the program may be seen as rather tricky and one may be not convinced
 about its correctness.   The author's description is rather operational.
So it should be useful to explain the program declaratively, and to provide
formal proof that it is correct.

In imperative and functional programming, program correctness implies that
the program produces the ``right'' results.  
In logic programming, which is nondeterministic, the situation is different. 
One also needs the program to be complete, i.e.\ to produce all
the results required by the specification.  (In particular, the empty program
producing no answers is correct whatever the specification is.)

This paper provides proofs of correctness and
completeness of the $n$ queens program; the proofs are declarative,
i.e.\ they abstract from any operational semantics. 

The paper is organized as follows.  After technical preliminaries, 
Section \ref{section.program} presents the $n$ queens program together 
with an informal description of its declarative semantics.
The next section presents a formal specification.
Proofs of correctness and completeness of the program 
are subjects of, respectively, 
Sections \ref{sec.correctness.proof} and \ref{sec.completeness.proof}.
The last section concludes the paper.
A version of this paper with slightly abridged initial sections has been
submitted to Theory and Practice of Logic Programming.

\section{Preliminaries}
\label{sec.preliminaries}
\paragraph{Basics.}
This paper considers definite clause logic programs.
We employ the standard terminology and notation \cite{Apt-Prolog},
and do not repeat here standard definitions and results.
We assume a fixed alphabet of function and predicate symbols.
The Herbrand universe will be denoted by \HU, the Herbrand base by \HB,
and the set of all terms (atoms) by \TU (respectively \TB\/);
$\HB_p$ is the set of ground atoms with the predicate symbol $p$.
By \NN we denote the set of natural numbers.
We sometimes do not distinguish a number $i\in\NN$ from its representation
as a term, $s^i(0)$. 
We use the list notation of Prolog. 
We assume that $[\seq e|e]$ stands for $e$ when $n=0$.
A list (respectively an open list) of
length $n\geq0$ is a
term $[\seq e]\in\TU$ ($[\seq e|v]\in\TU$), where $v$ is a
variable; $\seq e$ are the members of the (open) list.
We generalize the latter notion, and say that $e\in\TU$ is a
{\bf member} of a term $t\in\TU$ if $t=[\seq[k-1]e,e|e']$ (for some terms
$\seq[k-1]e,e'$, where $k>0$).
In such case we also say that $e$ is the $k$-th member of $t$.
Note that this kind of membership is defined by the Prolog built-in predicate
{\tt member}/2.

\comment{
   So what is a predicate for a logician, is a procedure for a programmer.

   So a predicate from the point of
   view of logic is a procedure from the point of view of a programmer.
}

Following \citeN{Apt-Prolog},
we use queries (conjunctions of atoms) instead of goals.
By an {\bf answer} of a program $P$ we mean any query $Q$ such
that $P\models Q$.  So an answer is a query to which a computed or correct
answer substitution has been applied; \citeN{Apt-Prolog} calls it computed/correct
instance of a query.
(Due to soundness and completeness of SLD-resolution, it does not matter
whether correct or computed answer substitutions are considered here.)
$\M_P$ stands for the least Herbrand model of a program $P$.
By the {\em relation} defined by a predicate $p$ in $P$ we mean
$
\Rel_P(p) =
\{\,\vec t\,\in\TU^n \mid P\models p(\vec t\,) \,\},
$
where $n$ is the arity of $p$.
\comment{%
By the {\em ground relation} defined by $p$ in $P$ we mean
$\Rel_P(p) \cap \HU^n$.
}
\comment{
(we may omit the subscript $P$ when this does not lead to ambiguity).
}

\paragraph{Specifications.}

In this paper,
the treatment of specifications and reasoning about correctness and
completeness follows that of \cite{Drabent.tocl16};
missing proofs and further explanations can be found there.
For further discussion, examples and references, see also
(\citeNP{Drabent.tplp18}; \citeNP{DBLP:journals/tplp/DrabentM05shorter}).

By a {\bf specification} we mean an Herbrand interpretation $S\subseteq\HB$.
A program $P$ is {\bf correct} w.r.t.\ a specification $S$ when
$\M_P\subseteq S$.  This implies that $S\models Q$ for any answer $Q$ of $P$.
A program $P$ is {\bf complete} w.r.t.\ $S$ when  $ S\subseteq\M_P$.
This implies that, for any ground query $Q$, if $S\models Q$ then $Q$
is an answer of $P$.
So $Q$ is an instance of an answer in each SLD-tree for $P$ and any query $Q_0$
more general than $Q$.

Dealing with the $n$ queens program we face a usual phenomenon:
Often it is inconvenient (and unnecessary) to specify $\M_P$ exactly, i.e.\ to
provide a 
specification $S$ for which the program is both correct and complete, 
$S=\M_P$.
It is useful to use instead an {\bf approximate specification}, which is a pair
$S_{\it c o m p l}, S_{corr}$ of specifications for, respectively,
completeness and correctness.
We say that  
a program $P$ is {\em fully correct} w.r.t.\ 
$S_{\it c o m p l}, S_{corr}$ when 
$S_{\it c o m p l}\subseteq\M_P\subseteq S_{corr}$.

\commenta{
 The choice of an approximate specification depends on the properties of
 interest. 
  See for instance
. . . 
  for various specifications for {\sc append}
  describing various properties of the program.
(. . . %
  is a version of this paper with extended introductory sections.)
}

 The choice of an approximate specification depends on the property of interest.
 As an example take the standard \APPEND program \cite{Apt-Prolog}.
 It does not define the list appending relation, but its certain
 superset  (as the program has answers with two arguments not being lists).
 To be sure that the first argument of $app$ is a list, it is sufficient to prove
 correctness of the program w.r.t.\
 $S_1=\{\, app(s,t,u)\in\HB \mid s \mbox{ is a list}\,\}$.
 Correctness w.r.t.\ 
 \[
 S_2=\left\{\, app(s,t,u)\in\HB \: \left|
 \begin{array}{l@{}}
                 \mbox{if } u=[\seq e] \mbox{ then for some } i\in\{0,n\}
                 \\
                 s=[\seq[i]e],\ t=[e_{i+1},\ldots,e_n]
 \end{array}
 \right.\right\}
 \]
 implies that \APPEND correctly splits a list given as the third argument 
 of $app$.
 (If the third argument is a list then the first two ones are a result of 
  splitting the list.)
 See \cite[(2) p.\,672 and Ex.\,3.4]{DBLP:journals/tplp/DrabentM05shorter}
 for two further specifications for correctness of \APPEND.
 The first one describes appending lists; the second, $S_{\rm APPEND}$,
 deals both with splitting and appending lists, and states that the first
 argument is a list.

 To be sure that the program will append any two lists, and split
 any list (of length $n$) in all possible ($n+1$) ways, we establish its
 completeness w.r.t.\ 
 \[
 S_{\it c o m p l} = 
 \left\{\,
  app( [\seq[m]e], [\seq {e'}], [\seq[m]e,\seq {e'}] ) \in\HB \mid m,n\in\NN
 \,\right\}.
 \]
 If we are interested only in lists of even length then we may consider
 completeness w.r.t.\ an appropriate subset of this specification.
 Without getting into details, we 
 note that the least Herbrand model $\M_{\rm APPEND}$ is distinct from all
 these specifications, and that \APPEND is complete w.r.t.\ $S_{\it c o m p l}$
 and correct w.r.t.\ $S_1, S_2$ and $S_{\rm APPEND}$.
 We have
 $S_{\it c o m p l} \subset \M_{\rm APPEND} \subset S_{\rm APPEND} \subset S_2
 \not\subset S_1
 $ 
 and $S_{\rm APPEND} \subset S_1 \not\subset S_2$.

For a discussion on building approximate specifications
see \hyperref[sec.comments]{\em Comments} at the end of
Section~\ref{section.program}.

%
%
%
%


\paragraph{Proving program correctness.}
An obvious way to prove correctness is to use the following sufficient
condition.
According to \citeN{DBLP:journals/tcs/Deransart93}, 
%
the condition is due to \citeN{Clark79}.

\begin{theorem}
\label{th.correctness}
    For a program $P$ and a specification $S$,
    if  $S\models P$  then  $P$ is correct w.r.t.\ $S$.
\end{theorem}
\vspace{-1.2\topsep}
\nopagebreak
\begin{proof}
 As $S$ is an Herbrand model of $P$, the least Herbrand model of $P$ is a
 subset of $S$.
\end{proof}

As $S$ is an Herbrand interpretation, \,$S\models P$ means that
     for each ground instance 
    \mbox{$    H\gets \seq B    $}
    ($n\geq0$)
    of a clause of $P$,
     if $\seq B\in S$ then $H\in S$.

So, informally speaking,
for correctness of a program it is sufficient that each its clause out of
correct atoms (i.e.\ those in the specification) produces only correct ones.

\paragraph{Proving program completeness.}
Informally, for completeness of a program w.r.t.\ $S$ it is necessary that
each atom $A\in S$ can be produced by some clause of $P$ (out of atoms
produced by $P$).
Moreover, $A$ should be produced in a finite way.
To formalize this idea we introduce some auxiliary notions.

\begin{definition}
  A ground atom $H$ is
  {\bf covered} {by a clause} $C$ w.r.t.\ a specification $S$
  if $H$ is the head of a ground instance  
  $
  H\gets \seq B
  $
  ($n\geq0$) of $C$, such that $\seq B\in S$
  \cite{Shapiro.book}.

  A ground atom $H$ is {\em covered by a program} $P$ w.r.t.\ $S$
  if it is covered w.r.t.\ $S$ by some clause $C\in P$.
\end{definition}

\begin{definition}
    A {\bf level mapping} is a function $|\ |\colon\HB\to\NN$.
    A program $P$ is {\bf recurrent} w.r.t.\ a level
    mapping $|\ |$ \cite{DBLP:journals/jlp/Bezem93} 
    when, for each ground instance $H\gets\seq B$ ($n\geq0$) of a clause of $P$
    and each  $i\in\{1,\ldots,n\}$, 
    we have $|H|>|B_i|$.
\end{definition}

The following sufficient condition is an immediate corollary 
of \cite[Theorem 5.6 and Proposition 5.4]{Drabent.tocl16}
or of \cite[Theorem 6.1]{Deransart.Maluszynski93},
and is sufficient for the purpose of this paper.
\comment{
 Th. 5.6 (semicompleteness), Proposition 5.4 (completeness)
}

\begin{lemma}
\label{lemma.completeness}
Let $P$ be a program, and $S$ a specification.  If 
each atom $A\in S$ is covered by $P$ w.r.t.\ $S$, and
$P$ is recurrent then $P$ is complete w.r.t.\ $S$.
\end{lemma}

\paragraph{A note on built-ins.}
\label{sec.built-ins}
The presented approach can be generalized in a rather obvious way to Prolog with
some built-ins.  We focus here on Prolog arithmetic.
A program $P$ using arithmetic predicates (like {\tt is}/2, or {\tt>}/2)
can be understood as augmented with an infinite set 
$P(Ar)\subseteq\HB$ of unit clauses
defining the ground instances of arithmetic relations \cite{Apt-Prolog}.
Such clauses are e.g.\,(in the infix form) 
$4\mathop{\tt is}2{+}2$, and 
$2{+}2\mathop{\mbox{\tt<}}7$.
To deal with correctness or completeness of such program, we assume that 
the specification is augmented with $P(Ar)$
(more precisely, the specification is
$S\cup P(Ar)$ where no arithmetic predicate occurs in $S$).
We also assume that $|B|=0$ for each $B\in P(Ar)$.
Now the sufficient conditions for correctness and completeness apply.
(As they are obviously satisfied by $P(Ar)$,
the condition for correctness
needs to be checked only for the clauses from $P$, and that for completeness
only for atoms from $S$.)

This approach abstracts from run-time errors.  So completeness w.r.t.\ $S$
means that
if $S\models Q$ and $Q$ is a ground instance of a query $Q_0$
then $Q$ is an instance of an answer of a Prolog computation starting with
$Q_0$, unless a run-time error or infinite loop is encountered.

\section{The $n$ queens program}
\label{section.program}

This section presents the $n$ queens program of
\citeN{Fruehwirth91}, and provides its informal declarative description.
Possible inaccuracies due to informal approach will be corrected in the next
sections, dealing with a formal specification and proofs.

The problem is to place $n$ queens on an $n\times n$ chessboard so that there
is exactly one queen in each row and each column, and at most one queen in
each diagonal. 
 The main idea of the program is to describe the placement of the queens by a
data structure in which it is impossible that two queens are placed on the
same row, column or a diagonal.%
\footnote{%
Actually, some of the diagonals are not dealt with.  This issue is clarified
later on.
}
In this way the constraints of the
problem are treated implicitly and efficiently.

This paper considers
the version of the program which represents natural numbers as terms in a
standard way.
Another version employs Prolog arithmetic.
The specifications and proofs of Sections
\ref{sec.spec} -- \ref{sec.completeness.proof} can be, in a rather obvious
way, transformed to ones dealing with the latter version,
following \hyperref[sec.built-ins]{\em A\,note\,on\,built-ins}\/ from the previous section.

Here is the main part of the program (with predicate names abbreviated);
it will be named \nqueens.%

\vspace{\abovedisplayskip}
\noindent
\mbox{}\hfill%
\begin{minipage}[t]{.8\textwidth}
{\small %
\begin{verbatim}
    pqs(0,_,_,_).
    pqs(s(I),Cs,Us,[_|Ds]):-
            pqs(I,Cs,[_|Us],Ds),
            pq(s(I),Cs,Us,Ds).

    % pq(Queen,Column,Updiagonal,Downdiagonal)  places a single queen
    pq(I,[I|_],[I|_],[I|_]).
    pq(I,[_|Cs],[_|Us],[_|Ds]):-
            pq(I,Cs,Us,Ds).
\end{verbatim}
}
\end{minipage}%
\hfill
 \begin{minipage}[t]{.035\textwidth}
\raggedleft
\small%
(\refstepcounter{equation}\theequation\label{clause1})
\\ \ \\
       \refstepcounter{equation}%
      (\theequation\label{clause2})%
\\ \ \\ \ \\ \ \\
       \refstepcounter{equation}%
      (\theequation\label{clause3})
\\[1.5ex]
       \refstepcounter{equation}%
        {(\theequation\label{clause4})}
\end{minipage}%
\vspace{\belowdisplayskip}
\\
Solutions to the $n$ queen problem are provided by those answers of \nqueens
that are of the form ${\it p q s}(n,q,t_1,t_2)$,
where $n$ is a number and $q$ a list of length $n$.
A number $i$ being the $j$-th member of list $q$ means that the queen
    of row $i$ is placed in column $j$.
(The role of $t_1,t_2$ will be explained later.)
    So to obtain the solutions, one can use a query
    ${\it p q s}(n,q_0,\myunderscore,\myunderscore)$,
    where $q_0$ is a list of $n$ distinct variables.

We quote the original description of the program, as it is an example of 
non declarative viewing of logic programs:
\begin{quotation}
  Observing that no two queens can be positioned on the same row, column
  or diagonals, we place only one queen on each row. Hence we can identify
  the queen by its row-number. Now imagine that the chess-board is divided
  into three layers, one that deals with attacks on columns and two for the
  diagonals going up and down respectively. We indicate that a field is
  attacked by a queen by putting the number of the queen there.

  Now we solve the problem by looking at one row at a time, placing one queen
  on the column and the two diagonal-layers. For the next row/queen we use
  the same column layer, to get the new up-diagonals we have to move the 
  layer one field up, for the down-diagonals we move the layer one field down.
\end{quotation}
This does not have much to do with the logic of the program; 
in particular the relations defined by the program are not described.
Instead, actions of the program are described
(and its data structures outlined).
Also, the description does not seem to justify why the program is correct.  
\comment{
 Fruehwirth spells up-diagonals, down-diagonals 
}
  Let us try to treat the program
declaratively, abstracting from the operational semantics.

\addtolength{\textfloatsep}{-1ex}
\addtolength{\textfloatsep}{-1ex}

\begin{figure}
\vspace*{-1ex}
\vspace*{-1ex}
\[
  \includegraphics
        [trim = 4.4cm 18.2cm 3.7cm 5.5cm, clip, scale=.95] 
  {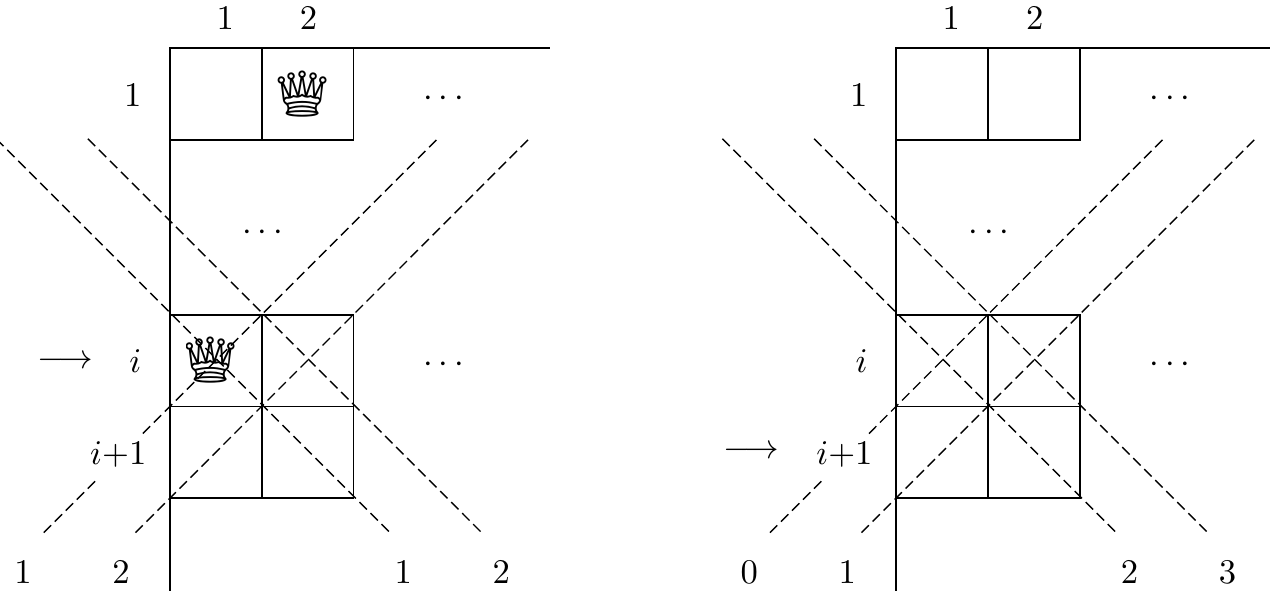}%
\vspace*{-1ex}
\]
  \caption[just a trick]{%
Numbering of rows and columns.  Numbering of diagonals in the
    context of row   $i$ (left), and $i+1$ (right).

The board with two queens is represented in the context of row $i$ as follows:
the columns by  $[i,1|\ldots]$, the up diagonals by  $[i|\ldots]$,
the down diagonals by $[i,\ldots,1|\ldots]$ (where $1$ is the member number
$i+1$).
Up diagonals with non-positive numbers are not represented. %
}
 \figrule
\label{figure.diagonals}
\end{figure}

\paragraph{Chessboard representation.}
Assume that columns and rows of the chessboard are numbered from the left/top.
Each queen is identified by its row number.
Diagonals intersecting a given row $i$ are numbered from the left
(Fig.\,\ref{figure.diagonals}).
In contrast to the numbering of rows and columns, 
this numbering is not fixed, it is 
specific to the context of the currently considered row;
the diagonal number $j$ includes the $j$-th field of the row.
Thus, in the context of row number $i$, its queen $i$ is
in the column and in the up and down diagonals of the same number.

To avoid ambiguity, let us state that 
by an up diagonal (resp.\ down diagonal) we mean the set of fields for which the
sum (the difference) of its row and column number is the same.
Given a set $A\subseteq\NN$ of queens, 
by a {\bf correct placement} of queens $A$ on a chessboard we mean one in which
each row, column, up diagonal and down diagonal contains at most one queen
from $A$.

When the initial query is as described above, the program 
represents a placement of queens by a list and two open lists.
Generally, the placement of queens on the columns, up diagonals
and down diagonals is represented by respectively three terms of the form
$[\seq[k]t|t]$.
If the column (respectively up diagonal, down diagonal) $j$, where $j>0$,
contains the queen $i$ then the $j$-th member of
the respective term 
is the number $i$.  
If it contains no queen 
then the $j$-th member of the term is arbitrary%
 \addtolength\footnotesep{-1ex}%
 \footnote{%
 It is a variable when the initial query is sufficiently general.
 }
or does not exist.
Such representation guarantees that at most one queen can be
placed in each column and each diagonal of a positive number.

\vspace{-.8ex}
\paragraph{Relations defined by \nqueens, rationale.}
Now we informally describe the purpose of the predicates of the program.
The role of ${\it p q}$ is to define  (a relation consisting of)
tuples $(i, cs, us, d s )\in\TU^4$,
where
\begin{equation}
\label{spec.attempt.pq}
\mbox{for some $j>0$, $i$ is the $j$-th member of each $cs,us,d s$.}
\vspace{-1ex}
\end{equation}
The role of ${\it p q s}$ is to define
tuples $(i, cs, us, [t|d s] )\in\NN\times\TU^3$
such that
\begin{equation}
\label{spec.attempt.pqs}
\begin{oldtabular}{l@{}}
 $i>0$ and
$cs$,\,$us$,\,$d s$ represent (as described above) 
\\
a correct placement of queens $1,\ldots,i$
on, respectively, \\
the columns, up diagonals and down diagonals,
\\
where the diagonals are
numbered in the context of row $i$,
\end{oldtabular}
\end{equation}
and additionally the tuples from $\{0\}\times\TU^3$.

Now we understand, for instance,
 why in clause (\ref{clause2}) the third argument of  
${\it p q s}$ is ${\it Us}$ in the head, and $[\myunderscore|{\it Us}]$ in
the body atom. 
 The latter represents up diagonals in the context of row $I$
(as $I$ is the first argument in the atom). 
Thus ${\it Us}$ represents them in the context of row $s(I)$ (which is the
first argument in the head).
In other words, if $[\myunderscore|{\it Us}]$ represents the up diagonals 
intersecting row $I$ then ${\it Us}$ represents those intersecting row $s(I)$.
Similar reasoning applies to the fourth argument and down diagonals.

Note that property (\ref{spec.attempt.pqs}) is not closed under substitution.
A triple $cs,us,d s$ that satisfies (\ref{spec.attempt.pqs}) may have instances
that do not (e.g.\ a list with a single occurrence of $i$ may have an
instance with multiple occurrences of $i$).  
So what we described differs from the relation actually defined by 
${\it p q s}$, and our description needs to be corrected.

\paragraph{Informal specification.}
\hspace{0pt plus .4em}
Note first that property (\ref{spec.attempt.pq}) is closed under
substitution (due to employing
the generalized notion of member).
Thanks to this our specification for ${\it p q}$ is obvious:
\begin{equation}
\label{spec.pq}
S_{p q} = 
    \{\,
      p q(\,i,\, [\seq[k]c,i|c],\, [\seq[k]u,i|u],\, [\seq[k]d,i|d] \,) 
      \in\HB   %
        {}\mid   k\geq0
      \,\}.
\end{equation}
Note that this specification is exact;
it is the set of atoms from $\M_{\rm NQUEENS}$ with the predicate symbol $p q$
(and it is the least Herbrand model of $\{(\ref{clause3}),(\ref{clause4})\}$).

\commenta{
  The difficulty in constructing a specification for ${\it p q s}$ 
  is that the program has
  also answers which represent incorrect
   placement of queens.
}  

We need to construct a specification for  ${\it p q s}$ 
so that correctness of \nqueens
implies that the program  solves the $n$ queens problem.
The difficulty is that the program has 
also answers which represent incorrect
 placement of queens.
%
 This cannot be avoided, as a non-ground triple of (open) lists $cs,us,d s$
 which represents a correct placement of queens may have instances which do not
 (e.g.\ because a single queen is placed in two columns). 
 It may seem that we face a contradictory task: the role of our specification
 is to describe correct placements, but it has to include some incorrect ones.

\commenta{
   The idea to overcome the difficulty is to care only about those atoms
   ${\it p q s(i, cs, us, d s )}\in\HB$, where $i>0$ and $cs$ is a list of distinct members.
   This leads to the following informal specification for correctness
   for ${\it p q s}$:
}

 To describe  ${\it p q s}$,
 note first that we are only interested in atoms 
 ${\it p q s(i, cs, us, d s )}\in\HB$, where $i>0$ and $cs$ is a list.
 The idea is that we care only about those in which the members of list $cs$ are
 distinct. 
 This means that the specification may {\em include} 
atoms ${\it p q s(i, cs, us, d s )}$  where $cs$
 is not a list, or is a list with a repeated member.  
 Whenever $cs$ is a list of distinct members, the remaining arguments are
 such that the whole atom describes a correct placement.
%
This leads to the following informal specification for correctness
for ${\it p q s}$:
\begin{equation}
\label{informal.spec.pqs}  
  \parbox{.77\textwidth}{%
    \hspace*{-1.5em}%
    the set of  those  ${\it p q s(i, cs, us, d s )}\in\HB$ where \\
    $i\in\NN$, \
    $1,\ldots,i$ are members of term  $c s$
    (cf.\ Section \ref{sec.preliminaries}, Basics), and\\
    if $i>0$ and $cs$ is a list of distinct members %
    then condition (\ref{spec.attempt.pqs}) holds.
}
\end{equation}
It follows that if ${\it cs}$ is a list of length $i$ then 
it is a list of distinct members $1,\ldots,i$, and hence
it is a solution of the $i$ queens problem.
Now our specification for \nqueens is the union of the sets described by
(\ref{informal.spec.pqs}) and (\ref{spec.pq}).
Note that it serves its purpose, as
correctness w.r.t.\ it implies that the program solves the problem
(each answer for the previously described query
 ${\it p q s}(n,q_0,\myunderscore,\myunderscore)$ represents a solution).
Note also that the specification is a proper superset of $\M_{\rm NQUEENS}$.

Such informal specification facilitates understanding of the program and
makes possible informal but precise reasoning
about the program.  
For an example, 
consider a ground instance of clause (\ref{clause2})
\[
\it p q s( s(i), cs, us, [t|d s]) \ \gets \
         p q s(i,cs,[t'|us],d s),\  p q(s(i),cs,us,d s).
\]
Assume that
the body atoms are compatible with the specification
(i.e.\ are in the sets (\ref{informal.spec.pqs}) and (\ref{spec.pq}),
respectively; we will refer to these two facts as
respectively ($\alpha$), ($\beta$)).
We show that also the head is compatible with the specification.
By ($\alpha$), 
$s(i) \in\NN$ and $1,\ldots,i$ are members of $c s$.
By ($\beta$), $s(i)$ is a member of $cs$.
Assume that $cs$ is a list of distinct members.
So by ($\alpha$) and (\ref{spec.attempt.pqs}),
 ${\it cs}$, ${\it[t'|{\it us}]}$ and the tail of ${\it d s}$
represent
a correct placement of queens $1,\ldots,i$ in the context of row $i$.
Hence
this placement in the context of row $i+1$ is represented by 
${\it cs, us, d s}$.
By ($\beta$)
 we have that, in the same context, $cs,us,d s$ represent
placing the queen $i+1$.
So its column (its up / down diagonal) is distinct from those occupied by
queens $1,\ldots,i$.
Thus ${\it cs, us, d s}$ represent a correct placement of queens $1,\ldots,i+1$.
Hence the head of the clause instance is in the set  (\ref{informal.spec.pqs}).

The reasoning of the last paragraph explains the clause and convinces us
about its correctness. 
Actually it is an informal outline of a central part
of a correctness proof of the program, based on Theorem \ref{th.correctness}.

\commenta{\vspace{-3ex}
\paragraph{Comment.}
    We exemplified a general pattern of constructing approximate
    specifications.  Some ground atoms may be irrelevant for the program
    properties we are interested in.  
    Thus they, or some of them, may be additionally included into the
    specification. 
    (In our case they are atoms ${\it p q s(i, cs, us, d s )}\in\HB$,
    where $i\in\NN$, $1,\ldots,i$ are members of $cs$, and moreover
     $i=0$ or $cs$ is not a list of distinct members.)
\comment{Submitted 2020.07:
    (In our case they are atoms ${\it p q s(i, cs, us, d s )}\in\HB$,
    where $i\in\NN$, $1,\ldots,i$ are members of $cs$, and
    $cs$ is not a list of distinct members or $i=0$.)
}
     So $\HB_{\it p q s}$
   is divided into a set $S_{\rm i r r}$ of irrelevant atoms and 
   $S_{\rm rel} = \HB_{\it p q s} \setminus S_{\rm i r r}$, the set of relevant
   ones.  From the latter a set $S_{\rm c}\subseteq S_{\rm rel}$ of
    ``correct'' ones is chosen,
   so that correctness w.r.t.\ specification $S_{\rm i r r}\cup S_{\rm c}$
   implies the program properties of interest.
   In many cases, $S_{\rm c}$ describes the ground atoms
   which the program should compute, so 
   $S_{\rm c}$ is used as a specification for completeness, and 
   $S_{\rm i r r}\cup S_{\rm c}$ as one for correctness.

   In the next section,
   the specification outlined here is made formal and is augmented by a
   specification for completeness.
}  

%
%
 %
 \paragraph{Comments.}
\label{sec.comments}
 Program \nqueens employs non-ground terms.  
 Given an initial query as described previously,
 the program uses open lists,
 and it seems crucial that the not yet assigned columns and diagonals are
 represented as unbound variables.   
 Our informal description above begins with a relation which is not closed
 under substitution.
 So one may suppose that the standard declarative semantics, based on the
 notion of logical consequence (and characterized by the least Herbrand models)
 is not suitable here.  Hence the notions of specification, correctness, and
 completeness of Section \ref{sec.preliminaries} would have not been suitable.
 One may expect that the s-semantics \cite{DBLP:journals/tcs/FalaschiLPM89}
 should be employed,
 as it makes it possible to explicitly deal with variables in program answers.

 Actually, this is unnecessary, as shown above and confirmed in the next
 sections.  We specify the program and prove its 
 correctness and completeness in terms of Herbrand interpretations.
 An initial version of this work
 was based on the
 s-semantics, this turned out to be less convenient and more complicated
 \cite{drabent2020arxiv.s-semantics}.

 The difficulty we face is to describe a certain property of non-ground atoms,
 by means of their ground instances, while an atom with the required
 property
 may have instances violating the property.
 The difficulty is overcome by simply neglecting such instances
  (and accepting them).
 This works, because a ground atom satisfying the property cannot be an
 instance of one not satisfying it, and each atom satisfying the property
 has a ground instance satisfying it.
 (The latter holds thanks to infinite \HU.)
 The neglecting is performed by describing in (\ref{informal.spec.pqs})
 the arguments $us,d s\in\HU$ of ${\it p q s}$ only for the cases in which
 $cs$ is a list of 
 distinct members.  Otherwise $us,d s$ are arbitrary.

 We applied a general way of constructing approximate specifications.
Some ground atoms may be irrelevant for the program 
properties we are interested in.  
So they, or some of them, may be additionally included into the specification.
The Herbrand base is split into a set $S_{\rm i r r}$ of irrelevant
(neglected) atoms,
and  $S_{\rm rel} = \HB \setminus S_{\rm i r r}$, the set of relevant
ones.
 And within $S_{\rm rel}$ we distinguish a subset 
$S_{\rm corr}\subseteq S_{\rm rel}$ of the answers we accept.
The specification for correctness is now $S = S_{\rm i r r}\cup S_{\rm corr}$.
Correctness w.r.t.\ $S$ implies the required properties of the program.
(Usually the intended initial queries have no instances in $S_{\rm i r r}$.)

In our case 
$S_{\rm corr}$ is the set of atoms describing correct placements of queens.
The irrelevant atoms are ${\it p q s(i, cs, us, d s )}\in\HB$,
     where $i\in\NN$, $1,\ldots,i$ are members of $cs$, and moreover
      $i=0$ or $cs$ is not a list of distinct members 
 ($us,d s$ are arbitrary).

 As may be expected, such approach
 does not lead to a unique specification.  
 The choice of $S_{\rm i r r}$ is not unique.
E.g.\ the first versions of this report used another specification
for the same property of interest.
   In the next section,
   the specification outlined here is made formal and is augmented by a
   specification for completeness.
%

\section{Approximate specification}
\label{sec.spec}
This section presents a pair of specifications %
for correctness and for completeness of \nqueens,
formalizing the ideas from the previous section.

The specification for predicate $p q$ is obvious.  
Both for correctness and for completeness it is
 $S_{p q}$
from (\ref{spec.pq}) in the previous section. 

In order to formulate the specification for ${\it p q s}$, we introduce some
additional notions. 
Let us first formalize the numbering of diagonals.
Assume a queen $j$ (i.e.\ the queen of row $j$) is placed in column $k$
(i.e.\ $j$ is the $k$-th 
member of a term $cs$ representing columns). 
Then, in the context of row $i$ (say $i\geq j$), the queen $j$ is on the up
diagonal of number $k+j-i$.
Similarly, the queen $j$ is on the down diagonal of number  $k+i-j$,
in the context of row $i$.
Consider, for instance, the queen $i-3$ placed in column 2.  Then,
in the context of row $i$,
it is on the up (down) diagonal number $-1$ (respectively $5$).

\begin{definition}
Let a queen (i.e.\ a number) $j$ be the $k$-th member of a list $cs$.

\quad
The {\bf up diagonal number} of $j$, w.r.t.\ $i$ in $cs$ is  $k+j-i$.

 \quad
The {\bf down diagonal number} of $j$, w.r.t.\ $i$ in $cs$ is  $k+i-j$.
\end{definition}

We can skip ``w.r.t.\ $i$'' when stating that some queens have distinct up
(down) diagonal numbers, as the numbers are distinct w.r.t.\ any $i$.
\pagebreak[3]

Now we are ready to introduce the core of our specification.

\begin{definition}
\label{def.correct.triple}
A triple of terms $(cs,us,d s)\in\TU^3$  
{\em represents a correct placement} up to row $m$ in the context of row $i$
(shortly: is {\bf correct} up to $m$ w.r.t.\ $i$\/) when $0\leq m\leq i$ and
\[
\qquad\ 
\begin{oldtabular}{l@{}}
  $cs$ is a 
list of distinct members, and each $j\in\{1,\ldots,m\}$ is its member,
\\
  \begin{oldtabular}{@{}l@{\quad}r@{}}
the up (respectively down) diagonal numbers of $1,\ldots,m$ in $cs$ are
distinct, 
&
\refstepcounter{equation}
(\theequation\label{correct.condition.distinct})
\\
          for  each  $j\in\{1,\ldots,m\}$,
          \\
  \quad
  \begin{oldtabular}{l}
      if
      the up (down) diagonal number of $j$ w.r.t.\ $i$ in $cs$ is $l>0$
          \\
      then the $l$-th member of $us$ (respectively $d s$) is $j$.
  \end{oldtabular}
  &
       \refstepcounter{equation}
      (\theequation\label{correct.condition1})
  \end{oldtabular}
\end{oldtabular}
\]
\end{definition}
Condition (\ref{correct.condition1}) assures that the placement of queens
$1,\ldots,m$ on the diagonals (according to $us,d s$) is compatible with
their placement on the columns, as described by $cs$.
Note that correctness of $(cs,us,d s)$ implies the 
required property of $cs$\/:\,
If  $(cs,us,d s)$ is correct up to $m$ then,
according to $cs$, the queens (of rows) $1,\ldots,m$ are placed
in distinct columns, up diagonals and down diagonals. 
(Note that the down diagonal numbers in (\ref{correct.condition1}) are
positive, as $j\leq i$.)

\pagebreak[3]

Now the specification for ${\it p q s}$ is
  \[
  \begin{array}[t]{@{}l@{}}
  S_{\it p q s} = 
     \begin{array}[t]{@{}l}
       \big\{\, {\it p q s}(0, cs, us, d s) \mid cs, us, d s \in \HU \,\big\}  
       \ \cup
       \\[1ex]
       \left\{\,  p q s(i, cs, us, [t|d s] ) \in\HB \: \left|\,
       \begin{oldtabular}{l@{\,}}
         $i>0$, \ \ \ %
               $1,\ldots,i$ are members of $cs$,
         \\
         if  $cs$ is a list of distinct members  then \\
          $(cs, us, d s)$ is correct up to $i$ w.r.t.\ $i$. \\
        \end{oldtabular}
        \right\}\right.,
     \end{array}
  \end{array}
\]
And our specification of \nqueens for correctness is
\[
S = S_{\it p q} \cup S_{\it p q s}.
\]

Note that correctness w.r.t.\ $S$ implies the required property of the program.
Take an atom $A = {\it p q s}(n, cs', us', d s')\in\TU$, such that $n>0$ and
$cs'$ is a list of length $n$.
If $S\models A$  then $cs'$ is a solution of the $n$ queens problem
(as, for each ground instance ${\it p q s}(n, cs, us, d s)$ of $A$,
$1,\ldots,n$ are members of $cs$,
 thus $cs$ is a list of distinct members $1,\ldots,n$, so
  $(cs, us, d s)$ is correct up to $n$ w.r.t.\ $n$.
Hence the up (down) diagonal numbers of  $1,\ldots,n$ in $cs$ are distinct,
so $cs$ represents a solution of the $n$ queens problem.

While specifying completeness, we are interested in ability of the program to
produce all solutions to the problem.  
This leads to the following specification for completeness:
\[
  S_{\it p q s}^0\ =\ 
   \begin{array}[t]{@{}l}
    \left\{\,  p q s(i, cs, us, [t|d s] ) \in\HB \: \left|
     \,
     \begin{oldtabular}{l@{\,}}
       $i>0$, \\
        $(cs, us, d s)$ is correct up to $i$ w.r.t.\ $i$. \\
      \end{oldtabular}
      \right.\right\}.  
   \end{array}
\]

We conclude this section with a property which will be used 
later on.
\pagebreak[3]
\begin{lemma}
\label{lemma.almost.equivalence}
Assume $0<m\leq i$.
Consider two conditions
\[
\begin{tabular}{c@{\qquad}l}
  $(cs,[t|us],d s)$ is correct up to $m$ w.r.t.\ $i$
   &  \refstepcounter{equation}  (\theequation\label{equivalence.part1})
\\
$(cs,us,[t'|d s])$ is correct up to $m$ w.r.t.\ $i+1$
   &  \refstepcounter{equation}  (\theequation\label{equivalence.part2})
\end{tabular}
  \]
For any $t,t'\in\HU$, (\ref{equivalence.part1}) implies
(\ref{equivalence.part2}). 
For any $t'\in\HU$, (\ref{equivalence.part2}) implies 
$\exists\, t\in\HU\:(\ref{equivalence.part1})$. 
\end{lemma}

\vspace{-1\topsep}
\begin{proof}
Assume that $cs$ is a list of distinct members and each $j\in\{1,\ldots,m\}$
is a member of $cs$.  We will consider here the diagonal numbers in $cs$.
Obviously, the up (down) diagonal numbers w.r.t.\ $i$ (of $1,\ldots,m$)
are distinct iff the diagonal numbers w.r.t.\ $i+1$ are.

Let $j\in\{1,\ldots,m\}$.
Then $l$ is the down diagonal number of $j$ w.r.t.\ $i$ 
iff
$l_1=l+1$ is the down diagonal number of $j$ w.r.t.\ $i+1$.
Note that $l>0$ (as $j\leq i$).
So for down diagonals,
 conditions (\ref{correct.condition1}) for $i$, $j$, $l$ and $d s$, and    
(\ref{correct.condition1}) for $i+1$, $j$, $l_1$ and $[t'|d s]$ are equivalent.

Number $l$ is the up diagonal number of $j$ w.r.t.\ $i$ 
iff
$l_2=l-1$ is the up diagonal number of $j$ w.r.t.\ $i+1$.
So $l_2\geq0$. For $l_2>0$ we, similarly as above, obtain that
for up diagonals conditions
\vspace{-1.5ex}
\vspace{-.5ex}
\begin{eqnarray}
    &&
    \label{lemma.almost.equivalence.cond.1}
    \mbox{(\ref{correct.condition1}) for $i$, $j$, $l$ and $[t|us]$} 
    \\
    &&
    \label{lemma.almost.equivalence.cond.2}
    \mbox{(\ref{correct.condition1}) for $i+1$, $j$, $l_2$ and $us$}
\end{eqnarray}
are equivalent.
For $l_2=0$, (\ref{lemma.almost.equivalence.cond.1}) vacuously implies
(\ref{lemma.almost.equivalence.cond.2}), and
(\ref{lemma.almost.equivalence.cond.2})
implies that (\ref{lemma.almost.equivalence.cond.1}) holds for some $t$,
namely $t=j$.

This completes the proof of both implications of the lemma.
\end{proof}

\section{Correctness proof}
\label{sec.correctness.proof}

Following Theorem \ref{th.correctness}, 
to prove correctness of program \nqueens w.r.t.\ specification $S$,
one has to show that $S$ is a model of each clause of the program.
In other words to show, for each ground
instance of a clause of the program,
that the head is in $S$ provided the body atoms are in $S$. 
For the unit clauses of \nqueens
\[
\begin{tabular}{l}
${\it p q(I,[I|\myunderscore],[I|\myunderscore],[I|\myunderscore])}$.
\\
${\it p q s}(0,\myunderscore,\myunderscore,\myunderscore)$.
\end{tabular}
\]
it is obvious that each ground instance of the clause is in $S$.
\
Consider clause (\ref{clause4}).
For any its ground instance
\[
{\it p q(i,[t_1|cs],[t_2|us],[t_3|d s])
\gets \it p q(i,cs,us,d s)
}.
\]
it immediately follows from the definition of  $S_{\it p q}$ that
 if the body atom is in $S$ (thus in $S_{\it p q}$) then its head is in 
$S_{\it p q}\subseteq S$.

The nontrivial part of the proof is to show that $S$ is a model of clause 
(\ref{clause2}).  Consider its ground instance
\[
{\it
    p q s(s(i),cs,{\it us},[t|d s]) \gets
            p q s(i,cs,[t_1|{\it us}],d s), \,
            p q(s(i),cs,{\it us},d s).
}
\]
Let $H$ be its head, and $B_1,B_2$ the body atoms.  Assume $B_1,B_2\in S$.
Now (by $B_2\in S$) 
$s(i)$ is the $l$-th member of  $cs, us, d s$ (for some $l>0$).
Note that $l$ is the up (down) diagonal number of $s(i)$ w.r.t.\ $s(i)$
in $cs$.
So condition (\ref{correct.condition1}) holds for  $s(i)$ w.r.t.\  $s(i)$.

Consider first the case of $i=0$.  Then
$(cs, us, d s)$ is correct up to $s(0)$ w.r.t.\ $s(0)$,
provided that $cs$ is a list of distinct members.
Hence $H\in S$.

Consider $i>0$. Note first that $1,\ldots,s(i)$ are members of $cs$
($s(i)$ as explained above, and $1,\ldots,i$ by $B_1\in S$).
Assume that
$cs$ is a list of distinct members.
Then (by $B_1\in S$)
$(cs,[t_1|{\it us}],d s')$ is correct up to $i$ w.r.t.\ $i$, where
$d s '$ is the tail of $d s$.
Hence by Lemma~\ref{lemma.almost.equivalence},
$\beta=(cs,{\it us},d s)$ is correct up to $i$ w.r.t.\ $s(i)$.
As shown above, 
(\ref{correct.condition1}) holds for 
$s(i)$ w.r.t.\ $s(i)$
(where $l$ is both the up and the down diagonal number of $s(i)$).
Thus (\ref{correct.condition1}) holds for $1,\ldots,s(i)$ w.r.t.\ $s(i)$.
Hence no up (or down) diagonal number of a $j\in\{1,\ldots,i\}$ is $l$.
As the latter diagonal numbers are distinct (due to $\beta$ being correct up to
$i$), 
(\ref{correct.condition.distinct}) holds for $1,\ldots,s(i)$.

  Hence $\beta$ is correct up to $s(i)$ w.r.t.\ $s(i)$.
  Thus $H\in S$.
  This completes the proof.

\section{Completeness proof}
\label{sec.completeness.proof}

As explained in Section \ref{sec.spec}, we are interested in completeness of
\nqueens w.r.t.\ specification 
$S_{\it p q s}^0$.
However
the sufficient condition of  Lemma \ref{lemma.completeness}
does not hold for this specification.
  Instead let us use 
\[
S^0 = S_{\it p q} \cup S_{\it p q s}^0 \cup 
     \{\, {\it p q s}(0, cs, us, d s) \mid cs, us, d s \in \HU \,\}  
\]
as the specification for completeness.%
\footnote{%
  This is a common phenomenon in mathematics;  an inductive proof of a property
  may be impossible, unless the property is strengthened.
  Actually, the same happened in the case of correctness. We are interested
  in correctness of \nqueens w.r.t.\ $S_{\it p q s}\cup\HB_{\it p q}$.
  However $S_{\it p q s}\cup\HB_{\it p q}$ is not a model of
  the program and Theorem \ref{th.correctness} is not applicable.  Instead
  we used a stronger specification $S = S_{\it p q s} \cup S_{\it p q}$.

  Obviously, correctness (completeness) w.r.t.\ a specification implies
  correctness  (completeness) w.r.t.\ any its superset (subset).
} %

We first show that each atom from specification $S^0$ is covered by program \nqueens.
Each atom
\vspace{-1.5ex}
\[
A =  p q(\,i,\, [\seq[k]c,i|c],\, [\seq[k]u,i|u],\, [\seq[k]d,i|d] \,) 
\]
from $S_{\it p q}$ is covered by \nqueens w.r.t.\  $S^0$; for $k=0$ by clause
(\ref{clause3}) as $A$ is its instance;
for $k>0$ by clause (\ref{clause4})
due to its instance 
\newcommand*{\SEQ}[3]
            {{\ensuremath{#1_{#2}, \allowbreak \ldots, \allowbreak #1_{#3}}}}%
$A\gets 
p q(\,i,\, [\SEQ c 2 k,\,i|c],\, [\SEQ u 2 k,i|u],\,\allowbreak
 [\SEQ d 2 k,i|d] \,) 
$ (as its body atom is in $S^0$).
Also, each atom $ {\it p q s}(0, cs, us, d s)$ is covered, as it is an
instance of  clause (\ref{clause1}).

The nontrivial part of the proof is to show that each $A\in S_{\it p q s}^0$
is covered.  Consider such atom, it is of the form
\vspace{-1.1ex}
\[
A = p q s(s(i), cs, us, [t|d s] ),  %
\]
where $i\geq0$ and $(cs,us,d s)$ is correct up to $s(i)$ w.r.t.\ $s(i)$.
So $cs$ is a list of distinct members, and each $j\in\{1,\ldots,s(i)\}$ is
a member of $cs$.
Let $s(i)$ be the $l$-th member of $cs$.  Thus
$l$ is the up and down diagonal number of $s(i)$ w.r.t.\ $s(i)$ in $cs$, and 
(by Def.\ \ref{def.correct.triple}) $s(i)$ is the $l$-th member of $us$ and
of $d s$.

We show that $A$ is covered by clause (\ref{clause2}) w.r.t.\ $S^0$, due to
its instance 
 \[
 A\gets  B_1, B_2. \qquad
 \mbox{where }
B_1 =  {\it p q s}(i,cs,[t'|us],d s),\ \
B_2 =    p q(s(i),cs,{\it us},d s)
 \]
(and $t'\in\HU$ will be determined later).
We have $B_2\in S^0$ (as $s(i)$ is the $l$-th member of $cs,us$ and $d s$).
If $i=0$ then $B_1\in S^0$, thus $A$ is covered by (\ref{clause2}).%
\footnote{%
    Note that in this case $A$ is covered  w.r.t.\ $S^0$ but not
    w.r.t.\ $S_{\it p q}^0\cup S_{\it p q s}^0$.
    This is why we use $S^0\supset S_{\it p q}^0\cup S_{\it p q s}^0$ as a
    specification. 
}

Assume $i>0$.
As $(cs, us, d s)$ is correct up to $s(i)$ w.r.t.\ $s(i)$, it is correct 
 up to $i$ w.r.t.\ $s(i)$, and 
by Lemma~\ref{lemma.almost.equivalence},
$(cs,[t'|us],d s')$ is correct  up to $i$ w.r.t.\ $i$, for some $t'\in\HU$, 
where $d s'$ is the tail of $d s$.
Hence for such $t'$ we have $B_1\in S_{\it p q s}^0\subseteq S^0$,
thus $A$ is covered by (\ref{clause2}).

This completes the proof that each $A\in S^0$ is covered by \nqueens
w.r.t.\ $S^0$.  
It remains to find
a level mapping under which \nqueens is recurrent.  Consider
the level mapping defined by
\vspace{-1ex}
\[
\begin{array}{l}
  | \, {\it p q s}(i,cs,us,d s) \,| = |i|+|cs|,
\\
  | \, {\it p q}(i,cs,us,d s) \,| = |cs|,
\end{array}
 \qquad \mbox{where} \qquad
    \begin{array}{l}
      |\, [h|t]\, | = 1+|t|,    \\
      |\, s(t)\, | = 1+|t|,    \\
      |f(\seq t)| = 0, 
    \end{array}
\]
for any ground terms $i,cs,us, d s, h,t,\seq t$, and any $n$-ary
 function symbol $f$ distinct from $s$ and from $[\ | \ ]$ ($n\geq0$). 
An easy inspection shows that
under this level mapping \nqueens is recurrent.
Hence by Lemma \ref{lemma.completeness}, the program is complete w.r.t.\ $S^0$.

\section{Conclusions}
 The paper provides an example of precise reasoning about the semantics of a
 logic
 program.  It presents detailed proofs of correctness and completeness of the $n$
 queens program of \citeN{Fruehwirth91}.  
 The program is short, but may be seen as tricky or non-obvious.
 The approach is declarative; the specifications and proofs abstract from any
 operational semantics, the program is treated solely as a set of logical
 formulae. 
 Note that
 in many cases, approaches based on the operational semantics are proposed 
 for reasoning about declarative properties of logic programs
  \cite{Apt-Prolog,DBLP:conf/tapsoft/BossiC89,DBLP:journals/jlp/PedreschiR99}.
  This seems to introduce unnecessary complications
  (cf.\ \cite[Section 3.2]{DBLP:journals/tplp/DrabentM05shorter}).

The original description \cite{Fruehwirth91} of the \nqueens program is rather
operational, as it mainly describes how the program works, and does not
explain the relations it defines
(cf.\ Section \ref{section.program}).
So we begin with informally describing the program declaratively, from a
logical point of view. 

The program uses non-ground data,
like open lists with some elements being variables.
Moreover some of its answers (which represent solutions to the problem)
have instances that do not.
So one may expect that approaches based on the standard semantics and
Herbrand interpretations are inapplicable here.  Actually, this is
not the case. 
We discuss difficulties with constructing a specification, show how to
overcome them, 
and provide a formal specification based on Herbrand interpretations.
Then we prove that the program is correct and complete with respect to the
specification. 

It may seem that s-semantics \cite{DBLP:journals/tcs/FalaschiLPM89}
is suitable here, as it explicitly deals with non-ground answers.
However the approach employed in this paper seems preferable, as
analogical specification and proofs
employing the s-semantics \cite{drabent2020arxiv.s-semantics} turn out to be
more complicated.

  Our specification is approximate;
 this means separate specifications for correctness and for completeness
 (see Section \ref{sec.preliminaries} for explanations and references). 
 Constructing an exact specification of the program would be too
 troublesome, and would result in more complicated correctness and completeness
 proofs.
 This is quite common in logic programming---one
 often does not need to know the exact semantics of one's program.  
  Some features of the program are of no interest, for instance they may be
  irrelevant to its intended usage.  So we do not need to describe them.
  What we require from a specification is that it describes those program
  properties in which we are interested.

This paper provides an example of precise declarative reasoning about the
semantics of a particular logic program.  
The case seems apparently difficult to deal with; we show how to overcome the
difficulties. 
In the author's opinion
the example confirms applicability of the employed approach
and provides hints for its use in other cases.

The detailed proofs presented here may be seen as too impractical due to
numerous details.  Note however that this is usually the case when proving
program properties.
  Experience with reasoning about programs,
  also in imperative programming,
  provides evidence that program correctness really does depend on many 
  details (see for instance the proof for Quicksort in
 \cite{AptBO.quicksort2009},
  and the example proofs in the papers mentioned above).
It should be possible to tame the complexity of proofs by employing 
some proof assistant.  This issue is however outside of the scope of this paper.
On the other hand,
  in the author's opinion proofs like those presented here
  can be performed by programmers at an informal level
  during actual programming,
at various degrees of precision.
Fragments of such informal reasoning (at two levels of precision) are shown in
Section \ref{section.program}.
We expect that formal proof methods, like those discussed here,  can teach
programmers a systematic way
 of reasoning about their programs in practice.

\paragraph{\bf Acknowledgement}
Comments of anonymous referees and of Michael Maher were instrumental in
improving the presentation.

\bibliographystyle{acmtrans}
\bibliography{bibpearl,bibmagic,bibs-s,bibshorter}

\end{document}